\documentclass[a4paper]{article}
\pdfoutput=1 
\usepackage{jcappub} 
\usepackage{amsmath,amssymb,bm,multirow}
\usepackage[dvipsnames]{xcolor}

\usepackage{siunitx}
\usepackage{relsize}
\usepackage{graphicx}
\usepackage[utf8]{inputenc}

\usepackage{mathtools}
\usepackage{algorithmic}

\usepackage{xcolor}
\usepackage[normalem]{ulem}

\begin{document}
\author[a,b]{Adrian E.~Bayer,}
\author[c,d,e]{Arka Banerjee}
\author[a,b]{and Yu Feng}

\affiliation[a]{Berkeley Center for Cosmological Physics, University of California,
Berkeley,  CA 94720, USA}
\affiliation[b]{Department of Physics, University of California,
Berkeley,  CA 94720, USA}
\affiliation[c]{Kavli Institute for Particle Astrophysics and Cosmology, Stanford University, 452 Lomita Mall, Stanford, CA 94305, USA}
\affiliation[d]{Department of Physics, Stanford University, 382 Via Pueblo Mall, Stanford, CA 94305, USA}
\affiliation[e]{SLAC National Accelerator Laboratory, 2575 Sand Hill Road, Menlo Park, CA 94025, USA}

\emailAdd{abayer@berkeley.edu}
\emailAdd{arkab@stanford.edu}
\emailAdd{yfeng1@berkeley.edu}


\title{
A fast particle-mesh simulation of 
non-linear 
cosmological structure formation with massive neutrinos
}


\abstract{
Quasi-N-body simulations, such as FastPM, provide a fast way to simulate cosmological structure formation, but have yet to adequately include the effects of massive neutrinos. We present a method to include neutrino particles in FastPM, enabling computation of the CDM and total matter power spectra to percent-level accuracy in the non-linear regime. The CDM-neutrino cross-power can also be computed at a sufficient accuracy to constrain cosmological observables. To avoid the shot noise that typically plagues neutrino particle simulations, we employ a quasi-random algorithm to sample the relevant Fermi-Dirac distribution when setting the initial neutrino thermal velocities. We additionally develop an effective distribution function to describe a set of non-degenerate neutrinos as a single particle to speed up non-degenerate simulations. The simulation is accurate for the full range of physical interest, $M_\nu \lesssim 0.6$eV, and applicable to redshifts $z\lesssim2$. Such accuracy can be achieved by initializing particles with the two-fluid approximation transfer functions (using the \textsc{reps} package). Convergence can be reached in $\sim 25$ steps, with a starting redshift of $z=99$. Probing progressively smaller scales only requires an increase in the number of CDM particles being simulated, while the number of neutrino particles can remain fixed at a value less than or similar to the number of CDM particles. In turn, the percentage increase in runtime-per-step due to neutrino particles is between $\sim 5-20\%$ for runs with $1024^3$ CDM particles, and decreases as the number of CDM particles is increased. The code has been made publicly available, providing an invaluable resource to produce fast predictions for cosmological surveys and studying reconstruction.
}

\maketitle


\section{Introduction}

Understanding the nature and generation mechanism of 
neutrino mass
is a challenge that unites particle physics with cosmology. Neutrino oscillation experiments were the first to provide evidence for neutrino mass \cite{SuperK, SNO, KamLAND, K2K, DayaBay}
by measuring the difference in the squares of the masses of the three neutrino mass eigenstates. The best fit results obtained from a joint analysis of oscillation experiments are $\Delta m_{21}^2 \equiv m_2^2 - m_1^2 \simeq 7.55 \times 10^{-5} {\rm eV}^2$ from solar neutrinos, and $|\Delta m_{31}^2| \equiv |m_3^2 - m_1^2| \simeq 2.50 \times 10^{-3} {\rm eV}^2$ from atmospheric neutrinos \cite{3sig_nu}. Because atmospheric neutrino experiments are only sensitive to the magnitude of the mass difference, there are two possibilities for the neutrino mass hierarchy: $\Delta m_{31}^2 > 0$, known as the normal hierarchy, or $\Delta m_{31}^2 < 0$, known as the inverted hierarchy. This leads to a lower bound on the sum of the neutrino masses, $M_\nu \equiv \sum_\nu m_\nu$, of $M_\nu \gtrsim 57 {\rm meV}$ for the normal hierarchy, or $M_\nu \gtrsim 96 {\rm meV}$ for the inverted hierarchy. An upper bound on neutrino mass, given by  $\beta$-decay experiments, is $M_\nu \lesssim 1.1 {\rm eV}$ \cite{Aker_2019}. 
While current particle physics experiments provide bounds, they are unable to determine either $M_\nu$, or the absolute mass scale of each eigenstate. 

By virtue of the high number density of neutrinos in the universe, cosmology provides 
a complementary probe to particle physics when studying various properties of neutrinos.
%
%
%
Numerous cosmological observables
can be used to study neutrino mass,
with one 
example being
the cosmic microwave background (CMB)
\cite{collaboration2018planck,choudhury2019updated, Haan_2016_SPT, CMB_Lensing, CMB_Lensing_2,Sherwin_2016_lens}, 
including secondary effects such as the thermal Sunyaev-Zeldovich (tSZ) effect \cite{Hill_2013_tSz,tSz} and kinetic Sunyaev-Zeldovich (kSZ) effect \cite{kSz,kSz2}. 
Another example is large-scale structure, which includes galaxy lensing, cosmic shear, and baryon acoustic oscillations (BAO) \cite{DES_Y1_gal, DES_Y1_Shear, Beutler_2014}. 
A further example is the Lyman-alpha forest \cite{Seljak_2005_Lya, Lya, Lya2_Palanque_Delabrouille_2015, Y_che_2017}. 
Assuming a $\Lambda{\rm CDM}$ cosmological model, the upper bound on the neutrino mass from the Planck 2018 CMB temperature and polarization data is $M_\nu < 0.26 {\rm eV}$ (95\% CL) \cite{collaboration2018planck}. Combining with BAO gives a more stringent bound of $M_\nu < 0.13 {\rm eV}$ (95\% CL), and further adding lensing gives $M_\nu < 0.12 {\rm eV}$ (95\% CL). Allowing more flexibility in the cosmological model, such as letting the spectral index 
run and considering a varying dark energy equation of state, can increase this upper bound to $M_\nu < 0.52 {\rm eV}$ (95\% CL) \cite{Valentino_2020}.
In all cases, the current upper bound on neutrino mass from cosmology is stronger than the bound from particle physics, and the cosmological bound is expected to improve with upcoming surveys.

This work focuses on the effects of neutrinos on cosmological structure formation  \cite{Doroshkevich_nu, Hu_1998, Eisenstein_1999, lesgourgues_mangano_miele_pastor_2013}.
Upcoming galaxy surveys, such as
DESI \cite{collaboration2016desi, Font_Ribera_2014}, 
LSST  \cite{LSST_Science_Book}, Euclid \cite{EuclidDefn, Euclid}, eBOSS \cite{eBOSS_Dawson_2016}, WFIRST \cite{spergel2013widefield}, and SKA \cite{SKA_Godfrey_2012, SKA_Villaescusa_Navarro_2015, SKA_Zhang_2020}, 
are predicted to give precise measurements of neutrino mass. For example, DESI and LSST forecast constraints of order  $\sim 0.02{\rm eV}$, 
thus the minimal neutrino mass should be detectable at the $\sim 3 \sigma$ level. Similar levels of accuracy are expected from CMB experiments when combined with BAO measurements from DESI.
%

Neutrinos affect structure formation because their low masses cause them to
behave as relativistic particles in the early universe, gradually becoming non-relativistic as the universe expands. This means that neutrinos possess high thermal velocities during the epoch of 
structure formation in the late universe, distinguishing them from the relatively slow cold dark matter (CDM).
As a result, massive neutrinos do not cluster on 
small scales,
leading to a relative suppression in the growth of matter perturbations 
compared to cosmologies with massless neutrinos. 
A useful way of quantifying the extent of small-scale suppression is to consider the ratio of the matter power spectrum between a cosmology with massive neutrinos, $P_m$, and a cosmology with massless neutrinos, $P_m^{M_\nu=0}$. To linear order, this is given by
\begin{equation}
    \frac{P_m}{P_m^{M_\nu=0}} \approx 1 - 8 f_\nu,
\end{equation}
where $f_\nu \equiv \Omega_{\nu,0}/\Omega_{m,0} \lesssim 0.05$ is the ratio of neutrino to total matter density at $z=0$ \cite{Hu_1998teg}. 
Hence, the relative suppression is proportional to the total neutrino energy density, which itself depends on $M_\nu$ as follows
\begin{equation}
    \Omega_{\nu,0} = \frac{M_\nu}{93.14 h^2 {\rm eV}}.
\end{equation}
Thus the matter power spectrum is sensitive to the sum of the neutrino masses.
Additionally, the profile of the matter power spectrum is, in principle, sensitive to the individual mass of each eigenstate \cite{LESGOURGUES_2006}; however, measuring the individual masses may not be achievable in the foreseeable future \cite{nu_individual_mass}. 

While the effects of neutrinos on linear (i.e.~relatively large) scales are well understood theoretically, understanding the effects on non-linear (i.e.~relatively small) scales is an active field of research.
For a fixed volume, there are many more independent modes on small scales
than on large scales.
Consequently, theoretical understanding of 
small scales would greatly increase the information that can be extracted from experimental surveys, in turn increasing the precision of their results. 
There is thus much motivation for simulations capable of modeling the effects of massive neutrinos on small scales. 

In recent years, many techniques have been developed to model structure formation with massive neutrinos, and they can mostly be separated into two methodologies.
The first is to use a fluid description for the neutrinos, and coupling this to the non-linear CDM gravitational potential \cite{Brandbyge_2009, Archidiacono_2016, Upadhye_2016, Ali_Ha_moud_2012, shoji2010massive, Inman_2017, senatore2017effective, Saito_2008, Dakin_2019}. Most of these approaches use linear theory, or perturbative approaches, to close the Boltzmann hierarchy in the absence of a known equation of state. 
These methods do not capture the full non-linear evolution of the neutrino field, leading to a reduction in accuracy at late times and on small scales, especially when the neutrino masses are much larger than the minimum total mass ($M_\nu \sim 57 {\rm meV}$).
The second methodology is to include neutrinos as an extra set of particles in the simulation \cite{Viel_2010, Bird_2012, Brandbyge_2010, Villaescusa_Navarro_2014, Costanzi_2013, Castorina_2014, Castorina_2015, Carbone_2016, Yu_2017,  Emberson_2017, Villaescusa_Navarro_2018, Adamek_2017, quijote}. 
This approach is fully non-linear, unlike the fluid approximation above.
Here the neutrino velocities are typically assigned by randomly sampling from the Fermi-Dirac distribution. This, however, can be problematic because the large neutrino thermal velocities cause a significant proportion of neutrino particles to traverse the simulation box multiple times, in turn erasing clustering on small scales and leading to shot noise.
The amount of shot noise in the neutrino power spectrum is inversely proportional to the number of neutrino particles, thus a simple approach to avoid this problem is by using a large number of particles. This is the approach taken in \cite{Emberson_2017}, but is extremely computationally expensive. 
A substantially faster approach is to use fluid-particle hybrid methods.
One example of a hybrid method is to use tracer particles to estimate the higher order moments of the Boltzmann hierarchy, requiring fewer particles \cite{Banerjee:2016zaa}. However, this method is still relatively expensive because it requires both neutrino particles and a non-linear neutrino fluid on a grid, which itself requires hydrodynamic techniques.
A more efficient hybrid method \cite{hybrid} treats fast neutrinos using a linear fluid approximation, while slow neutrinos are treated as particles after some user-defined redshift threshold. This minimizes computational cost, while still ensuring that the power induced by neutrino particle clustering is larger than the shot noise.

Recent work \cite{Banerjee_2018} provides a superior means to evade shot noise at all redshifts,
by sampling the Fermi-Dirac distribution in a low entropy, quasi-random manner\footnote{We use the term quasi-random to distinguish from pseudo-random, see e.g. \cite{SOBOL199055} for definitions of the two. Using a quasi-random sampling scheme ensures that the entropy of the underlying physical system is not increased by sampling.}. 
This has been shown to reduce the shot noise by a factor of $>10^7$,  enabling more accurate study of all scales. 
It is this method that we will focus on in this work.
A follow up study \cite{Brandbyge_2019}, in the context of hybrid simulations, showed that this method can induce spurious correlations between neutrinos and CDM on small scales. However, we note that this study did modify the methodology of \cite{Banerjee_2018} in a manner that could have exaggerated their results, by using a different sampling scheme for the Fermi-Dirac distribution and by initializing neutrino particles at a late redshift of $z = 4$ (as typical for hybrid methods).
We will show that such spurious effects can be avoided with simple considerations in our particle-only implementation.

While one could study neutrinos using a full N-body or hydrodynamic simulation, there is increased interest in quasi-N-body methods, such as FastPM \cite{fastpm} and COLA \cite{cola}, as they
produce significantly faster simulations of structure formation.
Some attempts have been made to include neutrinos in COLA by using fitting formulae for the growth factors \cite{Wright_2017}, but this approach does not provide the required accuracy in the non-linear regime for upcoming surveys. 
We focus on FastPM,
which implements a particle-mesh (PM) approach and enforces the correct linear 
evolution by using modified kick and drift factors.
Moreover, while quasi-N-body methods often fail at very small scales, typically $\lesssim 1{\rm Mpc}/h$, this has recently been addressed within FastPM by traversing the gravitational potential using gradient descent techniques
to increase the small scale force resolution \cite{Dai_2018}.
The purpose of this work is to add neutrino particles to FastPM by employing the methods of \cite{Banerjee_2018} to enable study of the non-linear regime. This will allow the inclusion of neutrino mass as one of the cosmological parameters in forward models, providing a fast way to interpret galaxy survey data.
Furthermore, after applying FastPM's inbuilt halo-finder, one can reconstruct the initial conditions of the universe\footnote{One example is BAO reconstruction, which was first considered in \cite{Eisenstein_2007_Recon} and applied in \cite{Padmanabhan_2012_ReconApplication}.} from galaxy positions and luminosities using the methods of \cite{Seljak_2017, Feng_2018, Modi_2018, Horowitz_2019}. Because massive neutrinos modify the information content of the CDM and total matter fields, this work will enable the study of massive neutrinos in the context of reconstruction.

It is worth mentioning another recently explored method to study massive neutrinos that uses neither the fluid approximation nor neutrino particles. This 
approach seeks to add the effects of massive neutrinos to the results of simulations without massive neutrinos, either by employing cosmological-rescaling algorithms \cite{Zennaro_2019}, convolutional neural networks \cite{giusarma2019learning}, or by perturbing the particles' final positions using a carefully designed gauge transformation \cite{partmann2020fast}.
We note that these methods ultimately rely on having full neutrino simulations available to compare against and tweak their input parameters, whereas our approach explicitly includes neutrinos using physical parameters as input.

The organization of this paper is as follows. In §\ref{sec:methodology} we outline the methodology employed to simulate massive neutrinos. We discuss the Fermi-Dirac sampling scheme, the setting of initial perturbations, and comment on changes made to FastPM's evolution algorithm. We then present the results of our simulation in §\ref{sec:results}, comparing with full N-body simulations such as Quijote \cite{quijote}. Comments on the runtime are given in §\ref{sec:runtime}. Finally, we conclude in §\ref{sec:conclusion}, outlining ideas for future work and applications to surveys.

\section{Methodology}
\label{sec:methodology}

This section outlines the approach used to include massive neutrinos within FastPM\footnote{ The code can be found at \url{https://github.com/fastpm/fastpm}. In the code, massive neutrinos are labelled as NCDM (not-cold dark matter), following the CLASS  \cite{class} convention.
}. We refer the reader to \cite{fastpm} for a comprehensive review of FastPM.

\subsection{Initializing massive neutrino particles}
\label{sec:FD}

\begin{figure}
\vspace{-0.8cm}
  \begin{center}
    \includegraphics[width=0.48\textwidth]{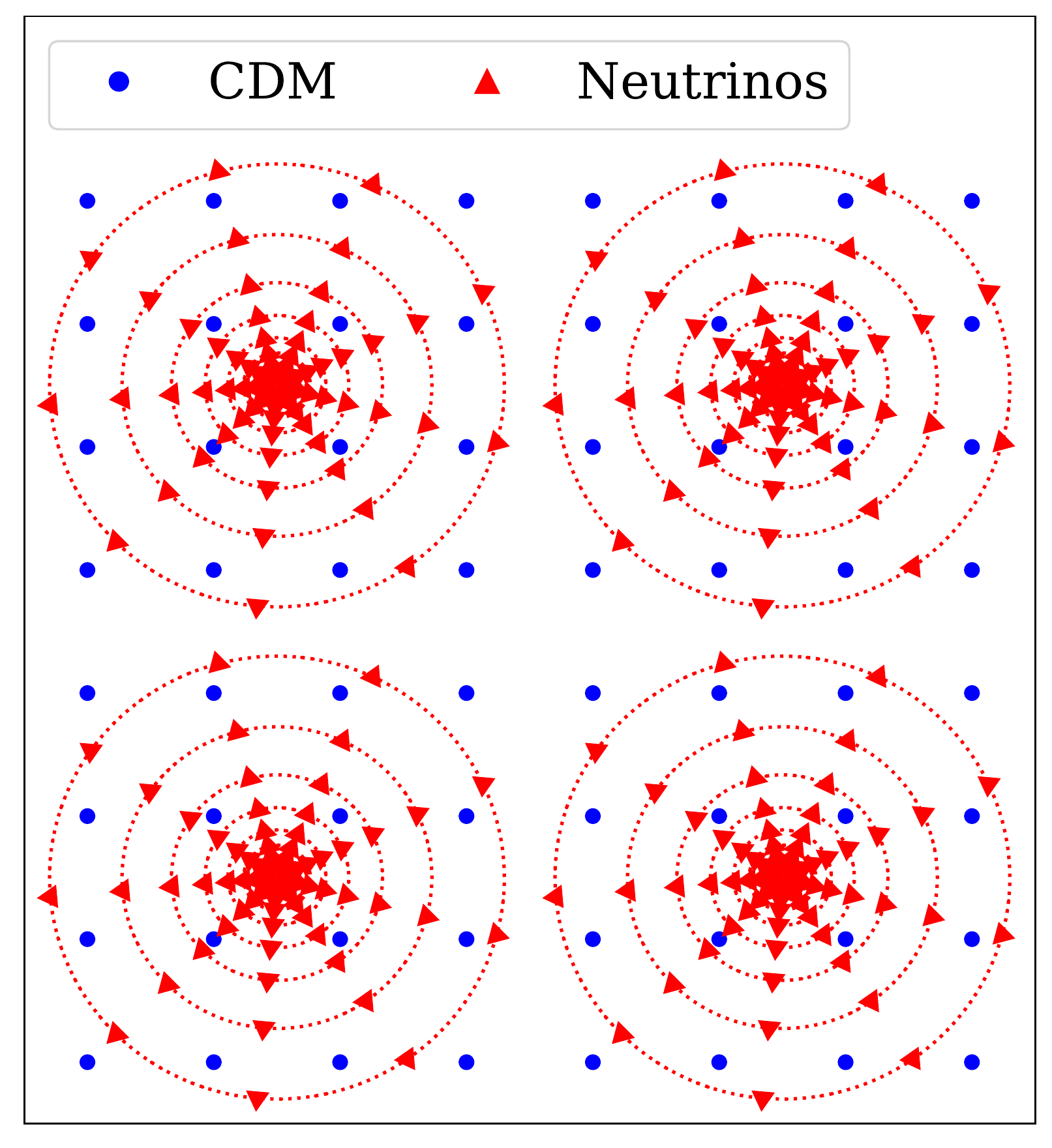}
  \end{center}
  \caption{2D illustration of the initial configuration for a neutrino grid 4 times coarser than the CDM grid. Each CDM particle is represented by a blue dot, and each neutrino particle is represented by a red triangle, with velocity in the direction of the triangle. Neutrino particles are initialized on spherical shells of radius proportional to the particle's speed. Each spherical shell is discretized according to a Fibonacci grid (not to scale).}
  \label{fig:IC}
\vspace{-0.5cm}
\end{figure}

To model the effects of massive neutrinos we include an additional species of particle in the simulation.
To set the initial thermal velocities of the neutrinos one must sample the Fermi-Dirac distribution. 
The sampling is usually performed randomly, leading to shot noise dominating small scales.
In this work we develop the methods of \cite{Banerjee_2018}, which have been shown to reduce this shot noise by a factor of more than $10^7$ by sampling the Fermi-Dirac distribution in a 
quasi-random manner.
This enables accurate study of small scales and late times.
We illustrate the initial configuration of  particles in Figure \ref{fig:IC}, and will describe the features of this setup throughout the remainder of this subsection.

A total of $N_n$ neutrino particles are initialized on a grid containing $N_{\rm sites}$ uniformly spaced sites. Each site comprises of $N_n / N_{\rm sites}$ particles, with each particle having a different initial thermal velocity.
Note this is different to CDM, which has $N_c$ grid sites and a single particle at each site.
The neutrino grid used in our applications is coarser than the CDM grid,
and we will show later that one can achieve accurate results using far fewer neutrino particles than CDM particles.

In dimensionless units, the Fermi-Dirac distribution function is given by
\begin{equation}
    f(q) = \frac{1}{e^q + 1},
    \label{fd}
\end{equation}
such that the number density of particles in an infinitesimal volume $d^3q$ centered at $q$ is given by
\begin{equation}
    dn = f(q) d^3q = q^2 f(q) dq d\Omega.
    \label{dn}
\end{equation}
In the final step we split the expression into the terms relevant for the velocity magnitude, $q^2 f(q) dq$, and direction (or solid angle), $d\Omega$.
To assign the velocity of each particle at a particular site, the magnitude and direction are assigned separately.
To assign the magnitude, the magnitude distribution $q^2 f(q)$ is discretely sampled at $N_{\rm shell}$ shells, such that the boundaries of the $i^{\rm th}$ shell are $(q_\mathrm{min}^{(i)}, q_\mathrm{max}^{(i)})$, where $1 \leq i \leq N_{\rm shell}$ . The left-most boundary is $q_\mathrm{min}^{(1)} = 0$, i.e.~zero velocity, and the right-most boundary $q_\mathrm{max}^{(N_{\rm shell})} \equiv q_\mathrm{max}$ is a numerical cutoff such that $q_\mathrm{max}^2 f(q_\mathrm{max})$ is negligible. The velocity magnitude of particles in the $i^{\rm th}$ shell is given by
\begin{align}
    q_{\rm shell}^{(i)} = \sqrt{ \frac{\int_{q_\mathrm{min}^{(i)}}^{q_\mathrm{max}^{(i)}} q^4 f(q) dq }{\int_{q_\mathrm{min}^{(i)}}^{q_\mathrm{max}^{(i)}} q^2 f(q) dq} },
    \label{q_shell}
\end{align}
and the mass of particles in the $i^{\rm th}$ shell is given by
\begin{align}
    m_{\rm shell}^{(i)} = \frac{\int_{q_\mathrm{min}^{(i)}}^{q_\mathrm{max}^{(i)}} q^2 f(q) dq }{\int_{0}^{q_\mathrm{max}} q^2 f(q) dq}.
    \label{m_shell}
\end{align}
The shell boundaries are chosen according to
\begin{align}
    \frac{1}{N_{\rm shell}} = \frac{\int_{q_\mathrm{min}^{(i)}}^{q_\mathrm{max}^{(i)}} g(q) dq }{\int_{0}^{q_\mathrm{max}} g(q) dq},
    \label{N_shell}
\end{align}
for some arbitrary kernel $g(q)$. This is designed so that each shell has an equal area under $g(q)$, with the choice of $g(q)$ 
depending on the application.
A natural choice would be to use the velocity magnitude distribution $g(q) = q^2 f(q)$, which splits the distribution into shells of equal phase-space volume. However, it was shown in \cite{Banerjee_2018} that better results are achieved by using $g(q) = q f(q)$, as this more finely samples the low-velocity tail of the distribution. This better resolves slow neutrino particles, which are most relevant for clustering. We thus employ this choice of kernel.

To study non-degenerate neutrinos one would have to separately sample the Fermi-Dirac distribution for each mass eigenstate and apply the above method multiple times. This would require more neutrino particles in the simulation and thus longer runtimes. To avoid this we use a single effective distribution to approximately describe all eigenstates. Imposing mass conservation, the appropriate distribution is given by
\begin{equation}
    \tilde{f}(q) = \sum_{j=1}^{N_\nu} \alpha_j^4 f(\alpha_j q),
    \label{eqn:FD_deg}
\end{equation}
where $\alpha_j \equiv m_j / m_1$ is the mass of eigenstate $j$ divided by the mass of eigenstate 1 (chosen to be the heaviest eigenstate), and $N_\nu$ is the number of eigenstates. This expression is exact when there are no cosmological perturbations. The derivation is included in appendix \ref{app:nondegFD}.

The velocity directions are chosen according to a Fibonacci grid \cite{fibonacci0, fibonacci1, fibonacci2} in which each shell isotropically emits particles in $2 N_{\rm fib} + 1$ directions, for integer $N_{\rm fib}$. We choose a Fibonacci prescription instead of the HEALPix \cite{healpix} implementation used in \cite{Banerjee_2018} as it gives more freedom when selecting the number of directions, and thus the number of neutrino particles. 
Hence, accounting for the discretization of the magnitude and direction, each neutrino site consists of $N_{\rm shell} \times (2 N_{\rm fib} + 1)$ particles. This gives the total number of neutrinos in the simulation,
\begin{equation}
    N_n =  N_{\rm sites} \times N_{\rm shell} \times (2 N_{\rm fib} + 1).
    \label{N_n}
\end{equation}

Because FastPM employs a Kick-Drift-Kick (KDK) algorithm \cite{quinn1997time}, as opposed to Drift-Kick-Drift, CDM particles initialized close to the neutrino sites will feel a gravitational attraction from the neutrinos and move towards the neutrino sites during the first kick of the simulation. The same is true for neutrino particles. This will produce spikes in the simulated power spectra at scales corresponding to the neutrino grid spacing, in a similar manner to the findings of  \cite{Brandbyge_2019}. To prevent such numerical artifacts, while keeping the more numerically stable KDK scheme, we take two precautions.
Firstly, the neutrino grid is staggered with respect to the CDM grid to separate the two species.
Secondly, the neutrino particles are initialized on spherical shells of radius proportional to their thermal velocity magnitude. The radii are chosen such that shells from different sites do not overlap.
Because neutrinos are orders of magnitude faster than CDM, 
this amounts to adding an infinitesimal drift step 
before the start of the KDK evolution.
This drift prevents large overdensities at the neutrino grid sites at the start of the simulation, in turn suppressing the spurious coupling caused
by particles getting drawn into neutrino grid sites.
We note that the effectiveness of these precautions is sensitive to the mass per neutrino particle. The more massive a neutrino particle, the stronger its gravitational pull on nearby particles. Thus for cosmologies with larger $M_\nu$, a larger value of $N_n$ is required to quell this effect. We will discuss the appropriate choice of $N_n$ in §\ref{sec:results}.

\subsection{Perturbations}
\label{sec:pert}
Having initialized the thermal velocities of the neutrino particles in the previous section, we must next include the effects of gravitational perturbations on the initial positions and velocities of all particles in the simulation. 
To do this, one would typically input the true
$z=0$ linear power spectrum from a Boltzmann solver such as CLASS \cite{class} or CAMB \cite{Lewis_2000}.
The simulation would then use a modified linear growth factor to backscale the power spectrum to the starting redshift of the simulation, and in turn set up the initial perturbations.
Because N-body simulations make various approximations, such as Newtonian dynamics, the growth factor used for backscaling is  modified to contain the same physics as the simulation's forward model. 
This is done to ensure that the results of the simulation on linear scales matches the true linear physics at $z=0$ \cite{Fidler_2017}. 
In the case of massive neutrino simulations, the forward model additionally includes both radiation and neutrinos, which must thus be accounted for when backscaling. This is a non-trivial procedure due to the scale-dependent growth introduced by massive neutrinos. We therefore perform backscaling using \textsc{reps} \cite{reps}, which applies the two-fluid approximation to compute the transfer functions of CDM and neutrinos. This is then used to obtain the power spectra and growth rates at the starting redshift of the simulation, and in turn compute the initial perturbations.
Moreover, 
while analytical forms for 
the 2LPT CDM and neutrino growth factors for massive-neutrino cosmologies have recently been presented in \cite{alej2020lagrangian}, there is currently no framework to apply this to generate non-linear initial conditions for simulations. 
We thus use the Zeldovich approximation when setting the initial perturbations, which requires starting the simulation at early times when non-linear effects are small ($z \gtrsim 99$).

In accordance with `scenario 4' of the \textsc{reps} paper \cite{reps}, we treat neutrino particles as non-relativistic in the forward model. Their mass is thus fixed throughout the evolution, and the total matter cosmological parameter used to source the gravitational potential (for example in Poisson's equation and the growth ODE of §\ref{sec:GrowthFactor}) is computed as 
\begin{equation}
    \Omega_m(a) = (\Omega_{c,0} + \Omega_{\nu,0}) a^{-3},
    \label{Om_pois}
\end{equation}    
where $\Omega_{c,0}$ and $\Omega_{\nu,0}$ are respectively the CDM and neutrino cosmological parameters at $z=0$.
The initial perturbations computed using \textsc{reps} are designed to account for this non-relativistic approximation and produce percent-level accuracy in the simulated power spectra at late times.
An advantage of \textsc{reps} over traditional backscaling is that \textsc{reps} is optimized to give agreement over a range of late redshifts, whereas traditional backscaling optimizes for a single redshift.
%
%
%
For consistency in this work, we will also use \textsc{reps} when initializing runs without massive neutrinos 
to enable comparison.
%

To compute the transfer functions, \textsc{reps} uses a Boltzmann solver, thus its output depends on the parameters used for the Boltzmann solver. In this work we modified the neutrino precision parameters in accordance with appendix B of \cite{Dakin_2019} to improve the accuracy of the transfer functions at small scales. In hindsight this was unnecessary as it causes little difference in the output of FastPM, so we plan to use the default neutrino precision settings in future work.

An alternative approach to backscaling would be to directly input the true linear power spectrum at the starting redshift together with the velocity transfer function. This method has been applied to small volume simulations \cite{hybrid}, but would require a more realistic forward model for accurate general implementation.

\subsection{Evolution}
\label{sec:GrowthFactor}

As outlined in section 2.4 of \cite{fastpm}, FastPM employs modified kick and drift factors to speed up convergence. This ensures the Zeldovich approximation is accurately followed at each timestep, using the Zeldovich equation of motion $x(a) = q + D(a) s$. 
We solve for the first order growth factor $D(a)$
using the following ODE \cite{peebles1993},
\begin{align}
    D''(a) +\left( 2 + \frac{E'(a)}{E(a)} \right) D'(a) = \frac{3}{2} \Omega_m(a) D(a), 
    \label{growthDE}
\end{align}
where 
$D' \equiv d D / d \ln a$,  $E(a) \equiv H(a)/H_0$ is the normalized Hubble parameter, and $\Omega_m(a)$ is given in equation \ref{Om_pois}. In this work, the background comprises of radiation ($\gamma$), CDM ($c$), neutrinos ($\nu$), and a cosmological constant ($\Lambda$), giving the appropriate Hubble parameter,
\begin{align}
    E(a) = \left[ \Omega_{\gamma,0}a^{-4} + \Omega_{c,0}a^{-3} + \Omega_{\nu}(a) E^2(a) + \Omega_\Lambda \right]^{1/2}.
    \label{Hubble}
\end{align}
The neutrino component is given by
\begin{align}
    \Omega_{\nu} (a) E^2(a) = \frac{15}{\pi^4} \Gamma_\nu^4 \frac{\Omega_{\gamma,0}}{a^4} \sum_{j=1}^{N_\nu} \mathcal{F} \left( \frac{m_j a}{k_B T_{\nu,0}} \right),
\end{align}
where $\Gamma_\nu \equiv T_{\nu,0}/T_{\gamma,0}$ is the neutrino-to-photon temperature ratio today, $m_j$ is the mass of neutrino species $j$, and 
\begin{align}
    \mathcal{F}(y) \equiv \int_0^\infty dx~ \frac{x^2 \sqrt{x^2+y^2}}{1+e^x}
\end{align}
is an integral arising from the Fermi-Dirac distribution while noting that neutrinos freeze out while relativistic \cite{reps}.
Note that the neutrino component of the background is treated exactly, with relativistic effects being accounted for.
On the other hand, neutrinos are treated as matter-like in the
source term of equation \ref{growthDE}, as motivated in §\ref{sec:pert}.
%

A further point of note is that equation \ref{growthDE} assumes the large-scale limit, i.e.~scales larger than the neutrino free-streaming scale. This neglects scale-dependent effects by treating neutrinos as non-relativistic particles, analogously to CDM.
Because the growth factor is just used by FastPM to speed up convergence, this limit is appropriate as it ensures accelerated convergence on large scales, while letting small scales converge naturally. 
We set the initial conditions to solve the ODE by assuming matter domination, 
giving
\begin{align}
    D (a_{\rm ini}) &= a_{\rm ini}, \\
    D' (a_{\rm ini}) &= a_{\rm ini}.
\end{align}
We use $z_{\rm ini} = 159$ to enable the simulation to begin at any time after this.

For users of FastPM, we note that FastPM previously assumed a $\Lambda$CDM background and thus employed the results of \cite{heath1977, peebles1980} to compute the growth factor, and  \cite{bouchet1994perturbative} to approximate the growth rate. This is unsuitable for neutrino simulations and has thus been replaced with the above.

\section{Results}
\label{sec:results}

\begin{figure}
    \includegraphics[width=\linewidth]{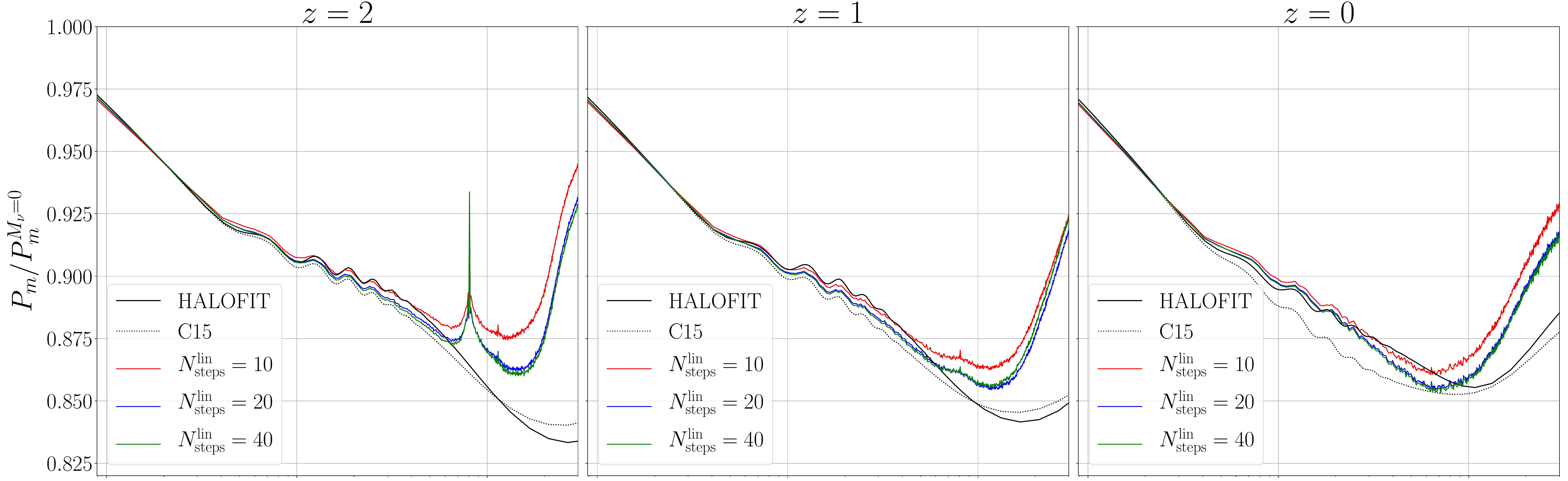}
    \includegraphics[width=\linewidth]{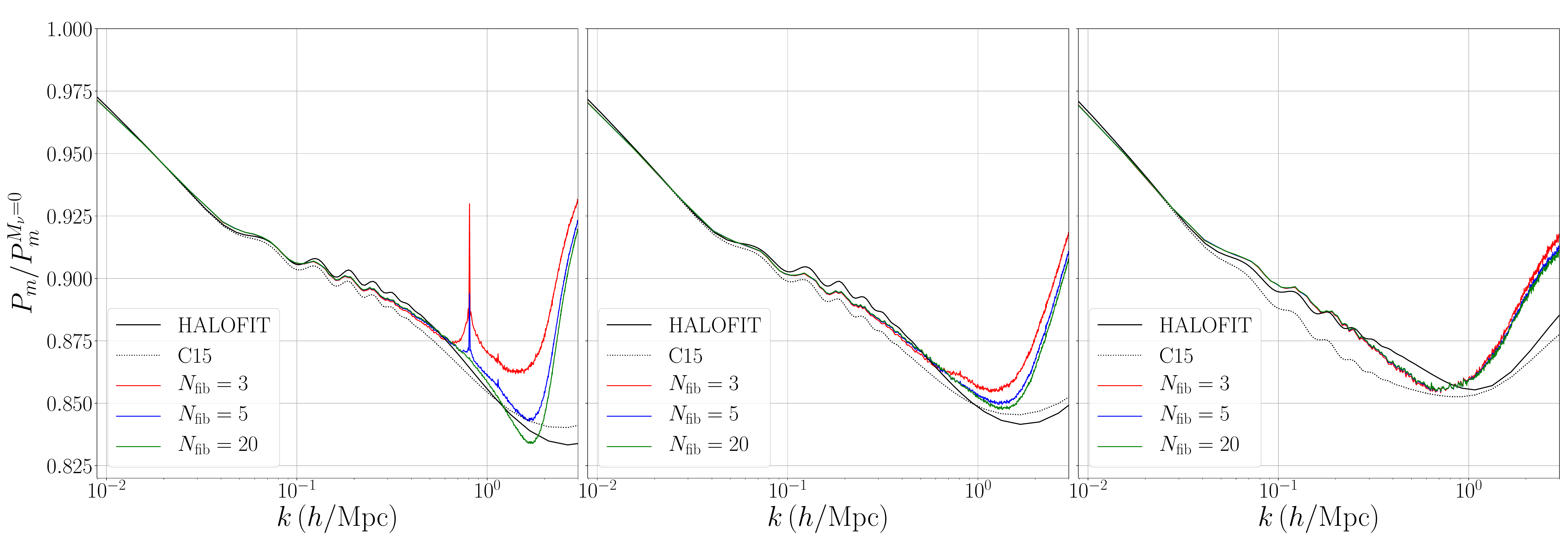}
  \caption{Ratio of massive to massless power spectrum at $z=2,1,0$ (left to right) for 3 degenerate $M_\nu = 0.2 {\rm eV}$ neutrinos.
  The top row shows the variation with $N_{\rm steps}^{\rm lin}$ for 
  fixed
  $N_{\rm fib} = 3$.
  The bottom row shows the variation with $N_{\rm fib}$ for 
  fixed $N_{\rm steps}^{\rm lin} = 20$.
  Theoretical predictions based on HALOFIT (solid black) and C15 \cite{Castorina_2015} (dotted black) are also shown.
  }
  \label{fig:MR}
\end{figure}

We consider a $1 {\rm Gpc}/h$ box with CDM and neutrino grid-numbers given by $N_c^{1/3} = 512$ and $N_{\rm sites}^{1/3} = 128$ respectively. The resolution of the force mesh is always chosen as $N_{\rm mesh}^{1/3} = 2 N_c^{1/3}$. The cosmological parameters are set as follows: $h = 0.6711$, $\Omega_m = 0.3175$, $T_\gamma = 2.7255 {\rm K}$, $N_{\rm eff} = 3.046$, 
$\Omega_k = 0$,
$A_s = 2.4 \times 10^{-9}$, and $n_s = 0.9624$. 
We begin by considering 3 neutrinos of total mass $M_\nu = 0.2 {\rm eV}$.
Simulations are started at $z=99$, at which time non-linear effects are small, as required for an accurate Zeldovich approximation. 
In order to achieve accurate results with a small number of timesteps, we first take 5 steps in $\log a$ until $z=19$, which is a sufficiently early time before non-linear neutrino effects come into play. We then take a further $N_{\rm steps}^{\rm lin}$ steps, spaced linearly in $a$, until $z=0$. Throughout this section we consider a single run of FastPM; averaging over many realizations would reduce variance, but is unnecessary for the purposes of this work. 

We first study the case of 3 degenerate neutrinos. Figure \ref{fig:MR} shows the ratio of the total matter (CDM+neutrinos) power spectrum between a cosmology with and without massive neutrinos. Different combinations of $N_{\rm steps}^{\rm lin}$ and $N_{\rm fib}$ are considered. 
For comparison, we also plot the theoretical predictions obtained from HALOFIT \cite{halofit0, halofit1, halofit_nu}, as well as the  modification of \cite{Castorina_2015} which will henceforth be referred to as C15. While these are not exact theoretical predictions they provide a useful diagnostic. 
%
%
Each column of Figure  \ref{fig:MR} represent a different redshift, $z=2,1,0$ from left to right. 
The top row considers the variation of $N_{\rm steps}^{\rm lin}$ while holding 
$N_{\rm fib} = 3$ fixed. It can be seen that all choices of steps produce accurate results on large scales, and that the result is suitably converged on small scales by $N_{\rm steps}^{\rm lin} = 20$. One important point to note is the occurrence of a spike at $z=2,1$ at $k \sim 0.8 h / {\rm Mpc}$. This spike corresponds to the spacing of the neutrino grid, and arises due particles being gravitationally attracted to the neutrino grid sites at the start of the simulation, as discussed in §\ref{sec:FD}.
This numerical artifact can be removed by distributing the neutrino mass over a larger number of particles, for example by increasing $N_{\rm fib}$.
To show this, the bottom row of Figure \ref{fig:MR} considers the variation of $N_{\rm fib}$ while holding
$N_{\rm steps}^{\rm lin} = 20$ fixed. It can be seen that $N_{\rm fib} = 5$ and $N_{\rm fib} = 20$ lead to accurate results at $z=1$ and $z=2$ respectively. Thus if one wishes to study these earlier redshifts, one must use the appropriate $N_{\rm fib}$. Table \ref{tab:params} summarizes some typical choices of parameters for runs with massive neutrinos. So far we have illustrated that \texttt{NC512\_NF3} is suitable for $z=0$ simulations, while \texttt{NC512\_NF20} should be used when one is interested in redshifts up to $z=2$. In the remainder of this section we will consider the \texttt{NC512\_NF20} run in order to study $z \leq 2$, unless stated otherwise.

\begin{table}[tbp]
    \centering
    \begin{tabular}{|l|r|r|r|r|r|r|r|r|}
    \hline
    Name & $N_c^{1/3}$ & $N_{\rm mesh}^{1/3}$ & $N_{\rm sites}^{1/3}$ & $N_{\rm shell}$ & $N_{\rm fib}$ & $N_{\rm steps}^{\rm log}$ & $N_{\rm steps}^{\rm lin}$ & Runtime Increase \\
    \hline
    \texttt{NC512\_NF3} & 512 & 1024 & 128 & 10 & 3 & 5 & 20 & 25\% \\
    \texttt{NC512\_NF20} & 512 & 1024 & 128 & 10 & 20 & 5 & 20 & 115\% \\
    \texttt{NC1024\_NF3} & 1024 & 2048 & 128 & 10 & 3 & 5 & 20 & 6\% \\
    \texttt{NC1024\_NF20} & 1024 & 2048 & 128 & 10 & 20 & 5 & 20 & 20\% \\
    \hline
    \end{tabular}
    \caption{\label{tab:params} 
    A summary of parameters used for some of the runs considered in this paper. Also included is the percentage increase in runtime due to massive neutrinos (discussed in §\ref{sec:runtime}). In all cases the force mesh is two times finer than the CDM grid. The two differences between the runs are the values of $N_c$ and $N_{\rm fib}$. Increasing $N_c$ enables studying smaller scales, while increasing $N_{\rm fib}$ enables studying higher redshift. As discussed in the text, $N_{\rm fib} = 3$ is suitable to study only $z=0$, but $N_{\rm fib} = 20$ is required for $z = 2$.}
\end{table}
 
\begin{figure}
    \includegraphics[width=\linewidth]{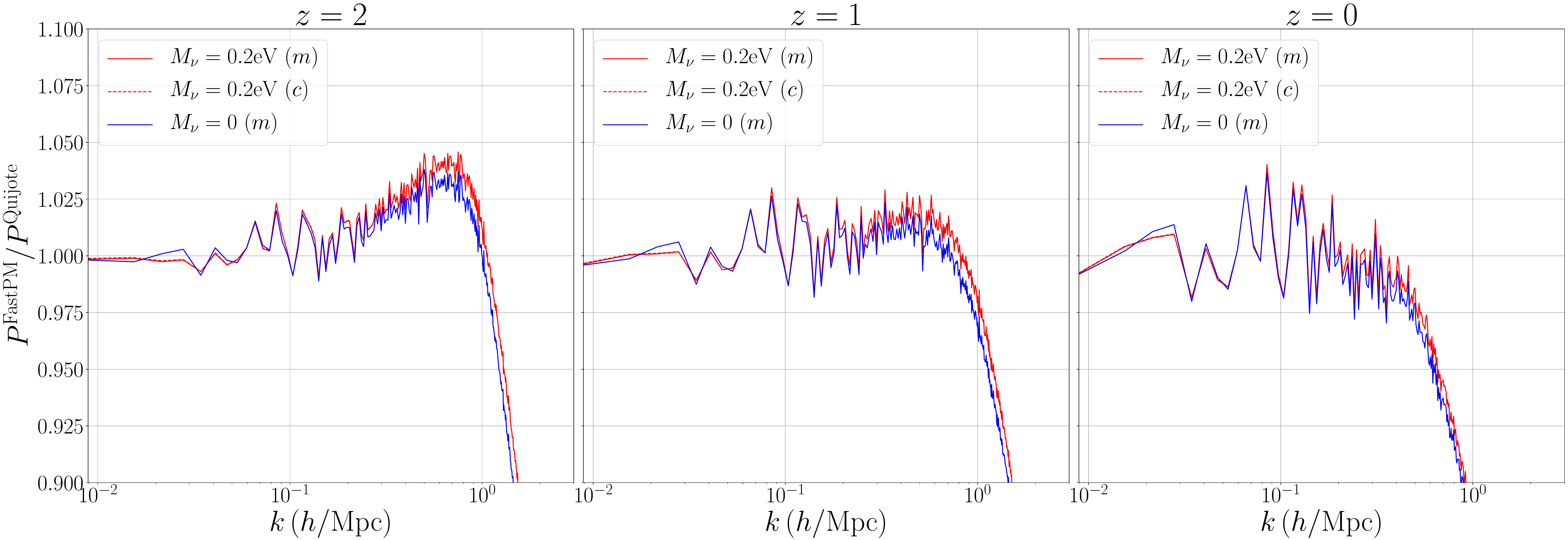}
    \caption{Comparison of the FastPM and Quijote power spectra at $z=2,1,0$ (left to right) for a cosmology with degenerate massive neutrinos with $M_\nu = 0.2 {\rm eV}$. Specifically, we consider the \texttt{NC512\_NF20} FastPM run and the ``$M_\nu^{++}$'' Quijote run. The solid red line compares the total matter power spectrum ($m$), while the dashed red line compares the CDM power spectrum ($c$) -- note it is difficult to distinguish the two by eye.
    Also included is
    a massless neutrino cosmology with matched $\sigma_8$ (blue).}
    \label{fig:pdm_quijote}
\end{figure}



\begin{figure}
  \includegraphics[width=\linewidth]{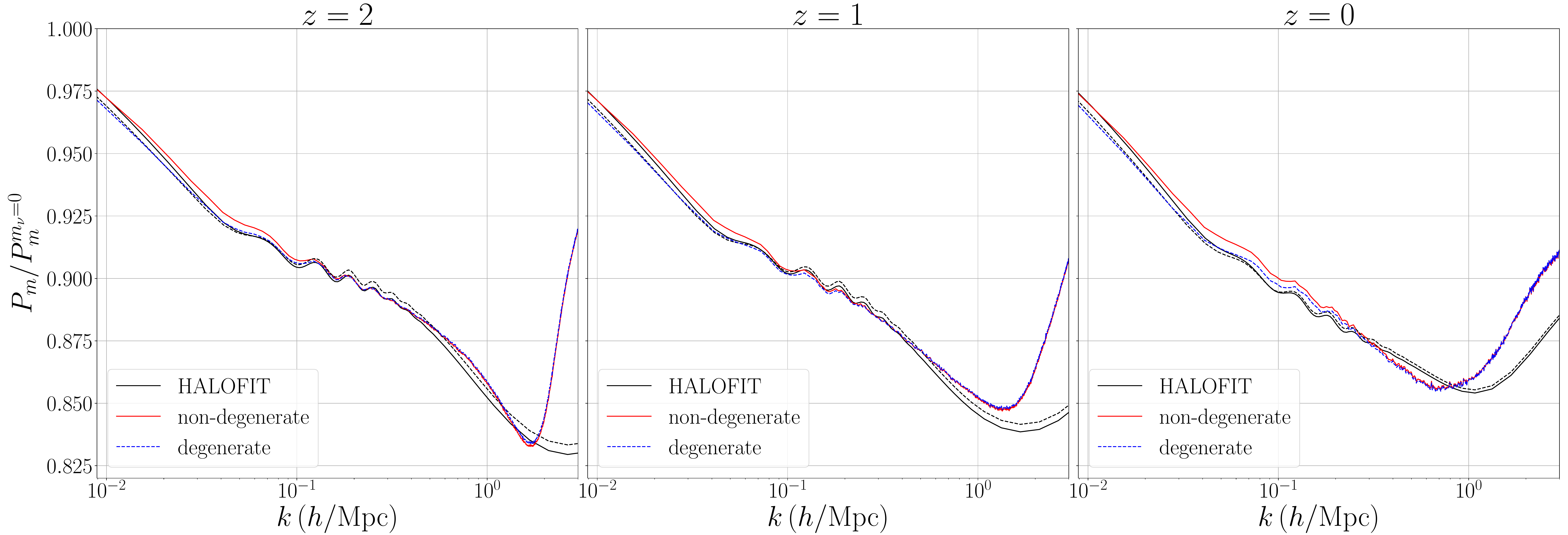}
  \caption{Ratio of massive to massless power spectrum at $z=2,1,0$ (left to right) for 3 non-degenerate (solid red) and degenerate (dashed blue) neutrinos with $M_\nu = 0.2 {\rm eV}$. The non-degenerate masses are ${0.12,0.06,0.02} \, {\rm eV}$. Theoretical predictions based on HALOFIT are shown for both the non-degenerate (solid black) and degenerate (dashed black) cases.}
  \label{fig:MR_nondeg}
\end{figure}

For a more careful analysis, Figure \ref{fig:pdm_quijote} compares the matter and CDM power spectra from FastPM with Quijote \cite{quijote}, a full N-body simulation. We consider the \texttt{NC512\_NF20} FastPM run and the ``$M_\nu^{++}$'' Quijote run.
For reference we also plot the matter power spectrum for a massless neutrino cosmology with matched $\sigma_8$. 
Firstly, it can be seen that both $P_c$ and $P_m$ show equally good agreement in the massive neutrino case -- the dashed red line overlaps the solid red line -- hence FastPM computes both power spectra with equivalent accuracy. Secondly, the difference between FastPM and Quijote is comparable in both the massive (red) and massless (blue) neutrino case, suggesting that any discrepancy with Quijote is not due to the inclusion of massive neutrino particles.
There is generally good agreement on large scales and an apparent under-prediction of the power on small scales. The reason for this is that while FastPM uses a particle-mesh approach to compute the forces throughout the simulation, Quijote employs tree methods at low redshift. This leads to Quijote producing more power on small scales, explaining the rapid drop in $P^{\rm FastPM}/P^{\rm Quijote}$ at large $k$ --- we note that this is not due to the shot noise present in Quijote. 
It can also be seen that there is a slight bump on intermediate scales at $z=2$, which is less prominent at lower redshift. We found that the bump grows when using a finer force mesh or initial-condition mesh in FastPM. We thus believe the bump is due to our use of a finer force mesh than that used by Quijote.
This has the effect of increasing the power on small scales, but is eventually dominated by Quijote's tree force calculation on small scales and late times, therefore it is only significant at $z=2$. The exact nature of the bump is also dependent on the parameters used in Quijote that define the redshift and scale at which the particle-mesh to tree transition occurs.

Next, we investigate the performance of our approximation for non-degenerate neutrinos given in equation \ref{eqn:FD_deg}. Figure \ref{fig:MR_nondeg} compares the massive to massless power spectrum ratio for the case of 3 neutrinos of masses $0.12, 0.06, 0.02$eV. 
The degenerate case is also included for reference. The agreement of the non-degenerate simulation with the theoretical lines is good on large scales and worsens on intermediate scales. This is likely because \textsc{reps} assumes degeneracy when computing the initial conditions, causing a relative gain in power. Even so, the non-degenerate results are of suitable accuracy for studying such mass schemes in the context of future surveys. 

\begin{figure}
  \includegraphics[width=\linewidth]{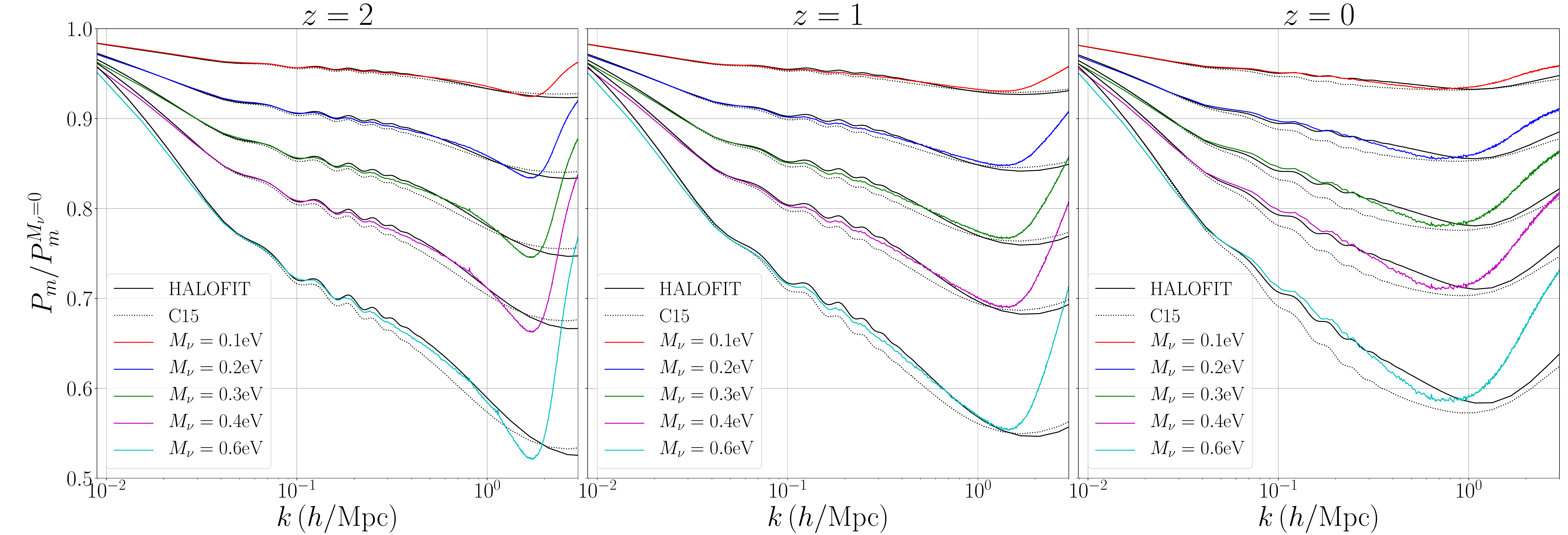}
  \caption{Ratio of massive to massless power spectrum at $z=2,1,0$ (left to right) for a variety of $M_\nu$, using $N_{\rm steps}^{\rm lin}=20$ and $N_{\rm fib}=20$. Note that 10 additional steps were taken at early $z$ for $M_\nu=0.6 {\rm eV}$, as discussed in the text. The theoretical predictions of HALOFIT (solid black) and C15 \cite{Castorina_2015} (dotted black) are also shown.}
  \label{fig:MR_mass}
\end{figure}

To investigate the accuracy of FastPM for different choices of neutrino mass, Figure \ref{fig:MR_mass} shows the ratio of the matter power spectrum between a massive and massless neutrino cosmology for a variety of choices of $M_\nu$. It can be seen that there is good agreement for the full range of physical interest ($M_\nu \lesssim 0.6 {\rm eV}$). 
Increasing $M_\nu$ beyond $0.2 {\rm eV}$ leads to a small spike at $z=2$ caused by the neutrino grid, as discussed in §\ref{sec:FD}. This is an expected result of the increase in mass per neutrino particle and can be alleviated by a small increase in $N_{\rm fib}$, or alternatively by increasing the number of steps at early redshift to prevent particles getting drawn into the neutrino grid sites. For $M_\nu = 0.6 {\rm eV}$, which is the upper bound of physical interest, the data in Figure \ref{fig:MR_mass} was generated using an extra 10 steps in $\log a$ between $z=99$ and $79$ to avoid the occurrence of a lager spike. 
While interest in cosmologies with $M_\nu = 0.6 {\rm eV}$ is limited, it is useful to know that accurate results can be achieved with an additional 10 steps compared to lower mass runs.

\begin{figure}
    \includegraphics[width=\linewidth]{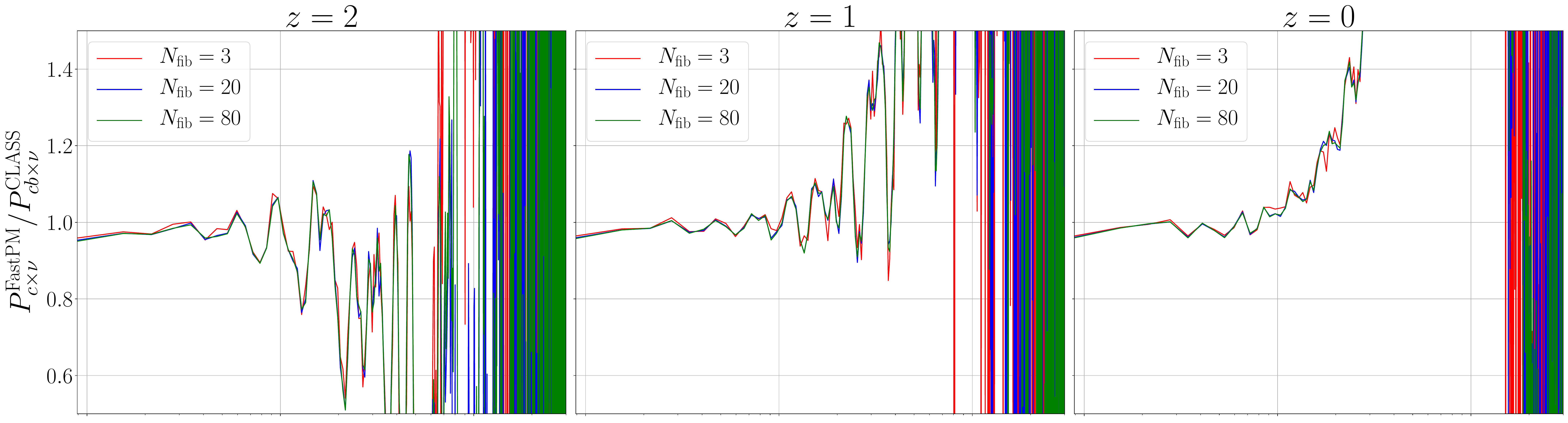}
  \caption{Comparison of the FastPM CDM-neutrino cross-power spectrum 
  with the linear cross-power from CLASS for $M_\nu=0.2 {\rm eV}$ using a variety of $N_{\rm fib}$. Note that, unlike FastPM, CLASS includes baryonic effects.
  }
  \label{fig:deg_nu}
\end{figure}

FastPM is also capable of computing the CDM-neutrino cross-power spectrum, as
required for observables such as galaxy-galaxy lensing. 
Figure \ref{fig:deg_nu} compares the FastPM cross-power to the linear cross-power computed by CLASS. There is good agreement on large scales, and the agreement worsens as $k$ increases due to non-linear effects that are not simulated by CLASS. There is negligible dependence on $N_{\rm fib}$ at large scales. We note that the cross-power is always weighted by a factor of $f_\nu$ in cosmological observables, thus one can tolerate larger error on the cross-power and still produce accurate observable predictions. 
We also found that the spikes discussed in §\ref{sec:FD} do occur in the cross-power at $z=2$, but are negligible for $z\leq1$. This effect can be reduced at $z=2$ by using a finer neutrino grid, which is relatively inexpensive for large $N_c$ simulations (as we will discuss in §\ref{sec:runtime}).

\begin{figure}
    \includegraphics[width=\linewidth]{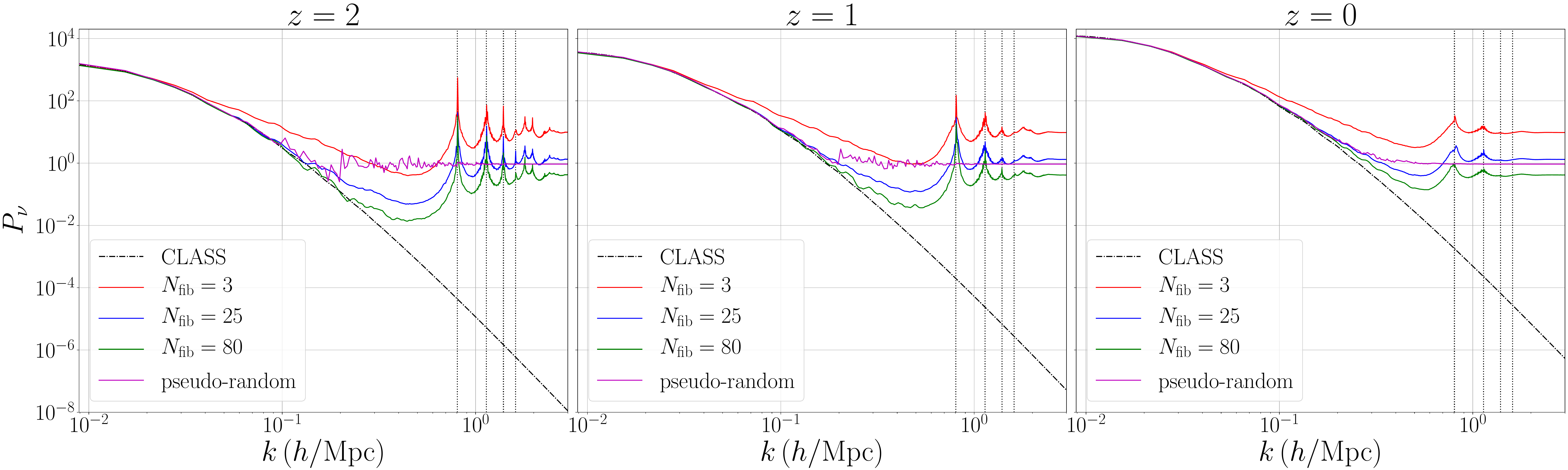}
  \caption{The FastPM neutrino power spectrum computed using a variety of $N_{\rm fib}$ for $M_\nu=0.2 {\rm eV}$. The linear result from CLASS is also included (black). For reference, a FastPM simulation initialized with pseudo-random neutrino thermal velocities is shown (magenta), and can be seen to produce shot noise. The total number of neutrinos, $N_n$, for the pseudo-random run is approximately the same as for the $N_{\rm fib}=25$ quasi-random run, hence the similar power as $k \rightarrow \infty$. Also shown are dotted vertical lines representing the four smallest wavenumbers associated with the neutrino grid: $k_n, \sqrt{2} k_n, \sqrt{3} k_n$, $2 k_n$.
  }
  \label{fig:Pn_shotnoise}
\end{figure}

While not directly observable, the neutrino power spectrum serves as a useful diagnostic for the quasi-random sampling scheme.
Figure \ref{fig:Pn_shotnoise} shows good agreement between the neutrino power spectrum computed by FastPM and CLASS on large scales. There is more sensitivity to $N_{\rm fib}$ compared to the cross-power, with larger $N_{\rm fib}$ required to ensure convergence at progressively smaller scales.
It can be seen that quasi-random sampling produces noisy $P_\nu$ on small scales, even with $N_{\rm fib} = 80$. To enable comparison with the noise produced by a typical pseudo-random sampling scheme we perform a FastPM simulation using pseudo-randomly sampled neutrino thermal velocities. 
As expected, a pseudo-random scheme produces shot noise, which in Figure \ref{fig:Pn_shotnoise} is manifested by the flattening of the power for $k \gtrsim 2 \times 10^{-1} h / {\rm Mpc}$.
For the pseudo-random example, a grid of $N_n^{\rm pseudo} = 1024^3$ neutrino particles was used, with one neutrino per grid site. Since shot noise is known to scale with the total number of neutrino particles, we compare with a quasi-random scheme using $N_{\rm sites} = 128^3$, $N_{\rm shell} = 10$ and $N_{\rm fib} = 25$, such that, using equation \ref{N_n}, $N_n/N_n^{\rm pseudo} = 128^3 \times 10 \times (2 \cdot 25 + 1) / 1024^3 = 0.996 \approx 1$. As expected, the $N_{\rm fib}=25$ and pseudo-random schemes approximately have the same power as $k \rightarrow \infty$ (there is a slight difference because the number of neutrino particles is not exactly matched). 
Firstly, it can be seen that the quasi-random scheme enables study of smaller scales compared to the pseudo-random scheme, and, ignoring spikes, has lower small-scale noise. Secondly, the difference in noise between the two sampling approaches becomes larger at earlier redshifts.
Thirdly, the pseudo-random power fluctuates around the more stable quasi-random power --- this can most clearly be seen at $z=2$ for $k \sim 10^{-1} h / {\rm Mpc}$. In fact, it was shown in \cite{Banerjee_2018} that such fluctuations are caused by early time artifacts produced by pseudo-random sampling, and also leaves a signature on $P_m$ at scales as large as $k \sim 10^{-2} h / {\rm Mpc}$.
Thus quasi-random sampling not only helps avoid shot noise on small scales, but also reduces noise at larger scales.

One apparent drawback of the quasi-random scheme is the introduction of spikes due to spurious correlations between CDM and neutrino particles.
Spurious correlations were first noted by \cite{Brandbyge_2019} in the context of hybrid simulations. However, this study did modify the methodology of \cite{Banerjee_2018} by using a different sampling scheme for the Fermi-Dirac distribution and by initializing neutrino particles at a late redshift of $z = 4$ (as typical for hybrid methods), making it difficult to present a direct comparison. To understand the nature of the spurious peaks observed in FastPM, we consider the Fourier transform of the initial neutrino grid. For a box of side length $1 {\rm Gpc}/h$ and $N_{\rm sites}^{1/3} = 128$ neutrino grid sites per side, the fundamental wavenumber is given by
\begin{equation}
    k_n = 2 \pi \frac{128}{1 {\rm Gpc}/h } \approx 0.8 h / {\rm Mpc}.
\end{equation}
This corresponds to the spacing between two adjacent neutrino grid sites. Because the grid is 3-dimensional, the next three smallest wavenumbers 
are $\sqrt{2} k_n$, $\sqrt{3} k_n$, and $2 k_n$. It can be seen in Figure \ref{fig:Pn_shotnoise} that the spikes in $P_\nu$ exactly align with these wavenumbers. This verifies the explanation in §\ref{sec:FD} that particles are drawn to the large overdensities at neutrino grid sites at the start of the simulation, in turn leaving a numerical artifact at late times. 
Hence, in our FastPM implementation, the spikes are 
a result of the coarse neutrino grid and can be removed by using a denser grid.
That said, $P_\nu$ has little effect on the small scale behaviour of cosmological observables as it is always weighted by a factor of $f_\nu^2$, and is itself small. Thus it typically suffices to use a coarse neutrino grid with a large enough number of neutrino particles per grid site,
as shown for $P_m$ in the discussion surrounding Figure \ref{fig:MR}.

\begin{figure}
  \begin{center}
    \includegraphics[width=\linewidth]{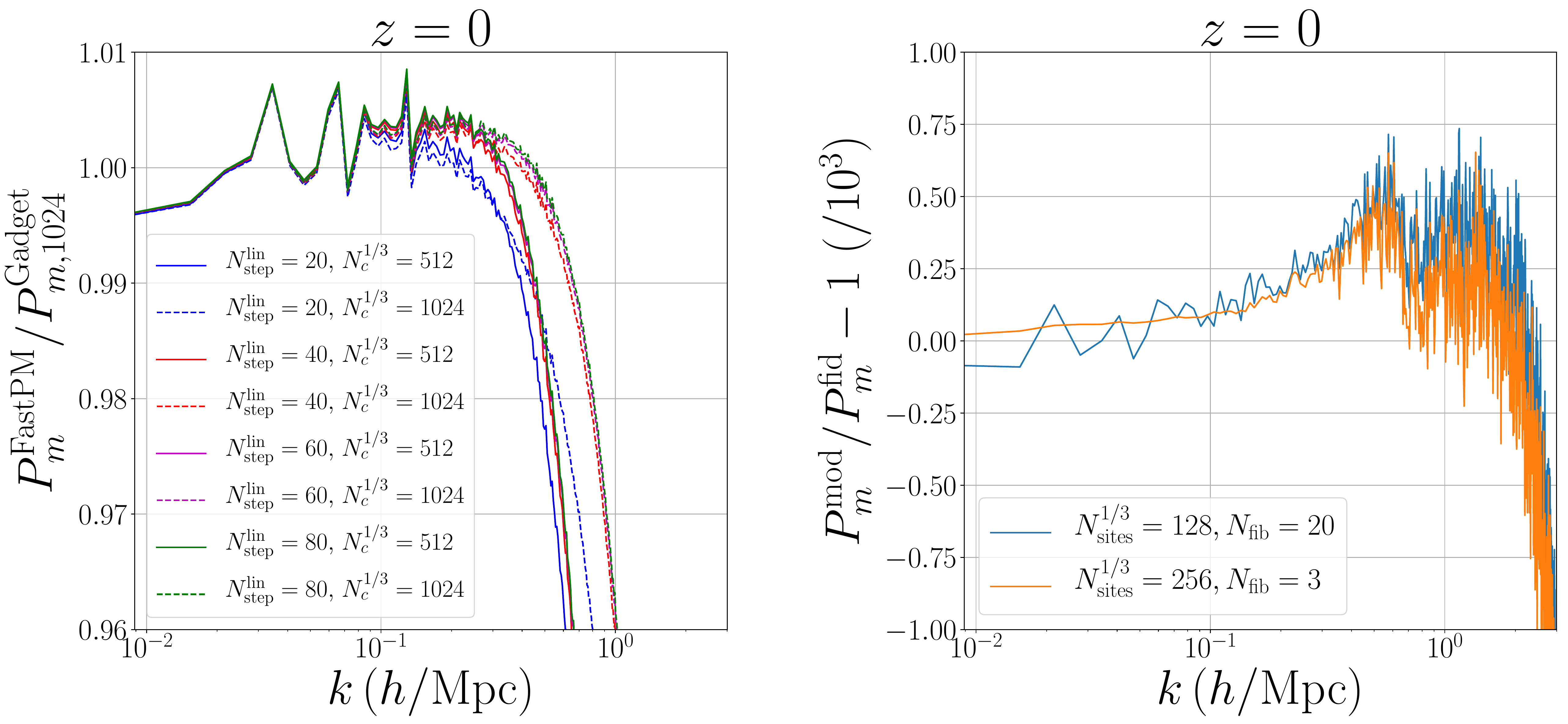}
  \end{center}
  \caption{
  (Left) Comparison of the FastPM matter power spectrum at $z=0$ for a run with degenerate massive neutrinos, $M_\nu = 0.12 {\rm eV}$, with an $N_c^{1/3} = 1024$ Gadget simulation. The step size and CDM grid are varied.
  (Right) Ratio of $P_m$ for FastPM runs with modified (mod) $N_{\rm sites}^{1/3}$ and $N_{\rm fib}$ compared to the fiducial (fid) values of 128 and 3 from \texttt{NC1024\_NF3} of Table \ref{tab:params}. Note that the ratio has been shifted by $-1$ and scaled by $10^3$, thus the vertical range represents a ratio range of $1 \pm 10^{-3}$, i.e.~$\pm 0.1\%$.
  }
  \label{fig:gadget}
\end{figure}

Finally, we consider using a finer CDM grid with $N_c^{1/3}=1024$. As a reference we use an $N_c^{1/3} = 1024$ Gadget \cite{Springel_2005} simulation for  degenerate neutrinos with $M_\nu = 0.12 {\rm eV}$.
The left of Figure \ref{fig:gadget} shows a fixed-amplitude comparison between the Gadget simulation and $N_c^{1/3} = 512$ \& 1024 FastPM simulations, considering a variety of step numbers. We use $N_{\rm fib}=3$ as we only compare $z=0$. In terms of Table \ref{tab:params}, we consider \texttt{NC512\_NF3} and \texttt{NC1024\_NF3}, while varying $N_{\rm steps}^{\rm lin}$. It can be seen that there is sub-percent agreement on large scales, and that using a two-times finer grid leads to agreement at approximately two-times higher $k$, as expected. Moreover, increasing $N_{\rm steps}^{\rm lin}$ to 40 extends the accuracy to slightly smaller scales, but the difference compared to $N_{\rm steps}^{\rm lin} = 20$ is small.

It is important to note that one does not need to increase $N_n$ with $N_c$ to obtain accurate results at small scales. This is illustrated on the right of Figure \ref{fig:gadget} where we consider modifications of \texttt{NC1024\_NF3} to increase $N_n$. It can be seen that increasing either $N_{\rm sites}^{1/3}$ or $N_{\rm fib}$ beyond the fiducial values of 128 and 3 causes a negligible ($<0.1 \%$) change in the $z=0$ power spectrum. This is a key finding in terms of studying smaller scales, as it shows one need only increase the number of CDM particles while keeping the number of neutrino particles fixed. This is aided by the fact that small scales are almost entirely dominated by the CDM evolution and the background cosmology.
While the results presented here are for a run with $N_c^{1/3} = 1024$, $M_\nu=0.12 {\rm eV}$, and $z=0$, we find similar results for larger $N_c$, $M_\nu$, and $z$ -- in all cases there is a sub-percent change in the power spectrum when increasing $N_n$.
Thus the relative increase in runtime-per-step caused by the inclusion of massive neutrinos will decrease as $N_c$ increases, enabling the study of small scales with only a small increase in runtime-per-step. We will now give a more thorough account of the runtime.

\section{Runtime}
\label{sec:runtime}

Firstly, we note that FastPM previously considered only CDM and $\Lambda$. To include massive neutrinos, this work has added functionality to simulate radiation in the background evolution by including photons and massless neutrinos. The runtime increase caused by this is negligible, thus it is the inclusion of neutrino particles, required to simulate massive neutrinos, that must be considered when studying runtime. To enable comparison in the following discussion, we compute the percentage difference in runtime-per-step between simulations of massive and massless neutrinos by using the same non-neutrino parameter values (including the number of timesteps) and number of CPUs in both cases.

Simulations were performed using the Cori supercomputer at the National Energy Research Scientific Computing Center (NERSC).
We first consider $N_{\rm fib}=3$ runs, shown in §\ref{sec:results} to give accurate results at $z=0$.
For 
the \texttt{NC512\_NF3} run of Table \ref{tab:params}, there are approximately equal numbers of CDM and neutrino particles: using equation \ref{N_n} gives $N_n/N_c= 10(2\times3 +1)/4^3 = 1.1$.
Such runs can be performed on a single Cori Haswell node in $\sim$ 715s, whereas the corresponding massless neutrino run takes $\sim$ 565s. Thus for this configuration there is a 25\% increase in runtime. 
We find that doubling both $N_c^{1/3}$ and $N_{\rm sites}^{1/3}$ requires 8 nodes and also shows a 25\% increase in runtime. 
However, as discussed at the end of the previous section, one does not need to increase $N_n$ as one increase $N_c$ -- accurate results can be achieved by increasing $N_c^{1/3}$ to 1024, while keeping $N_{\rm sites}^{1/3}=128$ fixed (\texttt{NC1024\_NF3}). Such a run requires only 4 nodes, and has an increase in runtime of 
6\%, as massive and massless runs take $\sim$1387s and 1307s respectively. As expected the change in runtime is sensitive to the ratio of $N_n$ to $N_c$, hence simulations with progressively larger $N_c$ and fixed $N_n$ have a smaller relative increase in runtime-per-step. This means that even runs with $N_{\rm fib}=20$, required for accurate results at $z=2$, only have an increase in runtime of 20\% for $N_c^{1/3}=1024$ (\texttt{NC1024\_NF20}).
Using a lower $N_c^{1/3}$ of $512$ with $N_{\rm fib}=20$ (\texttt{NC512\_NF20}) does lead to a larger runtime increase of 115\% because in this case $N_n/N_c \approx 6$. However, this large runtime increase is not problematic as $N_c^{1/3} = 512$ runs without massive neutrinos are relatively inexpensive anyway.
The key results of this paragraph are reported in Table \ref{tab:params}.

Massive neutrino runs typically require more timesteps than runs without massive neutrinos: while FastPM can achieve high accuracy for cosmologies without massive neutrinos in a couple of steps \cite{fastpm}, a massive neutrino simulation requires $\sim 25$ steps. This is because of the need to start simulations at an earlier redshift and to carefully capture the interplay between CDM and neutrinos, as documented in §\ref{sec:methodology}. Thus the increase in total runtime for massive neutrino simulations is dominated by the need for additional steps.
Note that we have ignored the effects of I/O and setting initial conditions; 
these scale with the total number of particles and will thus also lead to increases in runtime for large runs, but are typically subdominant for $N_c^{1/3} \lesssim 1024$.

\section{Conclusions}
\label{sec:conclusion}

This work has presented a fast and scalable particle-only method to study the effects of massive neutrinos on cosmological structure formation. This is enabled by three key ingredients. Firstly, we sample the Fermi-Dirac distribution in a low entropy, quasi-random manner when setting the neutrino initial conditions. This reduces the noise that typically plagues pseudo-random neutrino particle simulations, with the reduction becoming more significant at earlier redshift.
Secondly, while massive neutrinos introduce scale dependence, we use \textsc{reps} \cite{reps} to set the initial perturbations via the two-fluid approximation transfer functions. This allows us to treat neutrinos as non-relativistic particles throughout the evolution and achieve accurate results at low redshift. 
Finally, we incorporate the above methodology into FastPM \cite{fastpm}
to enable fast evolution. Altogether, the simulation 
produces accurate results for the matter, CDM, CDM-neutrino, and neutrino power spectra across the full range of neutrino masses permitted by current experimental constraints ($M_\nu \lesssim 0.6 {\rm eV}$) at $z \lesssim 2$. 
Furthermore, the increase in runtime-per-step due to massive neutrinos is small, as
the required number of neutrino particles is typically less than or similar to the number of CDM particles. Together with the fact a run requires $\sim25$ steps,
FastPM is considerably faster than alternative schemes based on full N-body simulations.

We have also addressed the problem of small scale spurious correlations caused by the quasi-random sampling method of \cite{Banerjee_2018}, found by \cite{Brandbyge_2019} for hybrid simulations. We have argued that, in our setup, spurious correlations are caused by the neutrino grid being coarser than the CDM grid, leading to nearby particles being attracted to the neutrino grid sites at the start of the simulation. Such spurious correlations in $P_m$ and $P_c$ can be adequately reduced 
by applying an infinitesimal drift step for neutrinos at the start of the simulation, and using a sufficiently large number of neutrino particles. 
The cross-power $P_{c \times \nu}$ 
is similarly
free of numerical artifacts arising from the sampling scheme --- this is true
at $z=2$ provided a sufficiently fine neutrino grid is used. 
Any remaining artifacts in $P_\nu$ are rendered subdominant by two effects: first, the contribution of the $P_\nu$ term to any observable is weighed by a factor of $f_\nu^2$, and second, $P_\nu$ itself is extremely damped on small scales compared to $P_c$. 

There are many avenues for future work. Our technique provides a quick way to predict the clustering of both CDM and total matter down to $\sim 1 {\rm Mpc}/h$ in the presence of massive neutrinos. Combined with FastPM's inbuilt halo finder \cite{fastpm}, analysis pipelines for fitting cosmological parameters can be built by interfacing with \texttt{nbodykit} \cite{Hand_2018}. This will enable the
prediction of galaxy-clustering and weak-lensing measurements for surveys such as DES \cite{DES_Y1_gal, DES_Y1_Shear}; one could implement an emulator-like approach \cite{Coyote_III, Kwan_2015, EuclidEmu, DeRose_2019, Zhai_2019, Wibking_2019, super-res_emu} 
to study the effects of massive neutrinos on clustering.
Moreover, recent work in effective field theory applied to BOSS \cite{Alam_2017_BOSS} has 
suggested that combining the full-shape BOSS data with Planck \cite{collaboration2018planck} can reduce the upper limit of the sum of neutrino mass to $M_\nu < 0.16 {\rm eV}$ (95\% CL) \cite{d_Amico_2020_EFT, Ivanov_2020}. 
One could test these results by performing a re-analysis of BOSS that considers small scale neutrino effects.
Finally, 
because massive neutrinos modify the information content of the CDM and total matter fields, one can use the techniques of \cite{Seljak_2017, Feng_2018, Modi_2018, Horowitz_2019} to study the effect of massive neutrinos on reconstruction.

\acknowledgments

We thank Uro\v{s} Seljak for fruitful discussion on the project. We also thank Yin Li and Patrick McDonald for insightful discussion on the methodology.
We acknowledge the use of \texttt{nbodykit} \cite{Hand_2018} for computing the power spectra presented in this work.
This research made use of the Cori supercomputer at the National Energy Research Scientific Computing Center (NERSC), a U.S.~Department of Energy Office of Science User Facility operated under Contract No.~DE-AC02-05CH11231.

\appendix
\section*{Appendices}
\section{Effective distribution for non-degenerate neutrinos}
\label{app:nondegFD}

We seek to describe a set of non-degenerate neutrinos with masses $\{ m_j \}_{j=1}^{N_\nu}$ by a single effective particle, for use in the sampling scheme described in §\ref{sec:FD}. Because equations \ref{q_shell}, \ref{m_shell}, and \ref{N_shell} are fractions of moments of the Fermi-Dirac distribution $f(q)$, we need only find the distribution function for the effective particle $\tilde{f}(q)$ up to a constant factor and arbitrary transformation of the argument.
Working in dimensionless units, as in equations \ref{fd} and \ref{dn}, the number of particles of eigenstate $j$ in an infinitesimal volume of size $d^3q_j$ is 
\begin{align}
    dn_j = f(q_j) d^3q_j.
\end{align}
The $j$ dependence arises due to the implicit dependence of $q_j$ on the non-degenerate mass $m_j$. Using the non-relativistic dispersion relation, the scaling of $q_j$ is given by $q_j \sim m_j$. We thus change variables to $q \equiv q_j/\alpha_j$ with $\alpha_j \equiv m_j / \mu$, for some constant with units of mass $\mu$, giving the number of particles of eigenstate $j$ in the \textit{common} infinitesimal volume $d^3q$,
\begin{align}
    dn_j =  f(\alpha_j q) \alpha_j^3 d^3q.
\end{align}
The number of effective particles in $d^3q$, denoted $d\tilde{n}$, is defined such that 
\begin{align}
    d\tilde{n} \equiv \tilde{f}(q) d^3q,
\end{align}
and the effective particle mass is denoted $\tilde{m}$. Enforcing mass conservation in each infinitesimal volume $d^3q$ gives
\begin{align}
    \tilde{m} d\tilde{n} &= \sum_j m_j dn_j \\
     \tilde{m} \tilde{f}(q) d^3q &= \sum_j m_j f(\alpha_j q) \alpha_j^3 d^3q.
\end{align}
Rearranging gives
\begin{align}
    \tilde{f}(q) = \frac{\mu}{\tilde{m}} \sum_j \alpha_j^4 f(\alpha_j q)
    \propto \sum_j \alpha_j^4 f(\alpha_j q),
\end{align}
which is the result stated in equation \ref{eqn:FD_deg}, having dropped the constant factor which is unneeded for the sampling algorithm.
The choice of $\mu$ to define $\alpha_j=m_j/\mu$ is arbitrary, but we choose $\mu=m_1$, the mass of the heaviest eigenstate, for two reasons. 
Firstly, as long as the mass ratios $\alpha_j$ are close to 1, a good sampling
for the heaviest eigenstate also implies a good sampling for the other mass eigenstates. If the mass ratio of an eigenstate is much smaller than 1, then there will be no significant clustering for this light eigenstate, and the sampling scheme is irrelevant.
Moreover, this choice ensures that $\tilde{f}$ will equal the correct Fermi-Dirac distribution in the degenerate limit, because $\alpha_j \rightarrow 1 ~\forall~ j$.

\bibliography{refs}
\bibliographystyle{JHEP}

\end{document}